\documentclass[11pt,a4paper]{article}
\pdfoutput=1
\usepackage{jcappub}

\usepackage{graphicx}
\usepackage{amssymb}
\usepackage{amsmath}
\usepackage{epsfig}
\usepackage{dcolumn}
\usepackage{rotating}
\newcommand{\ba}{\begin{eqnarray}}
\newcommand{\ea}{\end{eqnarray}}
\newcommand{\be}{\begin{equation}}
\newcommand{\ee}{\end{equation}}
\newcommand{\al}{\alpha}

\newcommand{\da}{\delta}
\newcommand{\la}{\lambda}
\newcommand{\za}{\zeta}

\newcommand{\oa}{\omega}

\newcommand{\La}{\Lambda}

\newcommand{\cF}{{\cal F}}

\newcommand{\+}{^{\dagger}}

\newcommand{\ra}{\rightarrow}
\newcommand{\Ra}{\Rightarrow}

\newcommand{\LF}{\left(}
\newcommand{\RF}{\right)}

\newcommand{\Rd}{\right.}

\newcommand{\mx}{\mbox}

\newcommand{\mand}{\mx{ and }}
\newcommand{\for}{\mx{ for }}

\newcommand{\with}{\mx{ with }}

\newcommand{\ie}{{\it i.e.\ }}

\newcommand{\pd}{\partial}
\newcommand{\D}{\nabla}

\newcommand{\Fc}{\mathcal{F}}
\newcommand{\Hc}{\mathcal{H}}

\newcommand{\Kc}{\mathcal{K}}

\newcommand{\Zc}{\mathcal{Z}}
\newcommand{\Wc}{\mathcal{W}}
\newcommand{\Pc}{\mathcal{P}}

\newcommand{\const}{\text{const}}

\title{Stable bounce and inflation in non-local higher derivative cosmology}
\author[a]{Tirthabir Biswas,}
\affiliation[a]{
Department of Physics, Loyola University, 6363 St. Charles Avenue, Campus Box 92, New Orleans, USA}
\emailAdd{tbiswas@loyno.edu}
\author[b]{Alexey S. Koshelev,}
\affiliation[b]{Theoretische Natuurkunde, Vrije Universiteit Brussel and The International Solvay Institutes,
Pleinlaan 2, B-1050,
Brussels, Belgium}
 \emailAdd{alexey.koshelev@vub.ac.be}
\author[c]{Anupam Mazumdar}
\affiliation[c]{Consortium for Fundamental Physics,
Physics Department,
Lancaster University,
Lancaster, LA1 4YB, UK and\\
Niels Bohr Institute,
Blegdamsvej-17,
Copenhagen-2100, Denmark}
\emailAdd{a.mazumdar@lancaster.ac.uk}
\author[d]{and Sergey Yu. Vernov}
\affiliation[d]{
Instituto de Ciencias del Espacio, Institut d'Estudis Espacials de Catalunya,
Campus UAB, Facultat de Ci\`encies, Torre C5-Parell-2a planta, E-08193,
Bellaterra (Barcelona), Spain and\\
Skobeltsyn Institute of Nuclear Physics, Lomonosov  Moscow State University,
Leninskie Gory 1, 119991,
Moscow, Russia}
\emailAdd{svernov@theory.sinp.msu.ru}

\abstract{
One of the greatest problems of primordial inflation is that the inflationary space-time is past-incomplete. This is mainly because Einstein's GR suffers from a space-like Big Bang singularity. It has recently been shown that ghost-free, non-local higher-derivative ultra-violet modifications of Einstein's gravity may be able to resolve the cosmological Big Bang singularity via a non-singular bounce. Within the framework of such non-local cosmological models, we are going to study both sub- and super-Hubble perturbations around an inflationary trajectory which is preceded by the Big Bounce in the past, and demonstrate that the inflationary trajectory has an ultra-violet completion and that perturbations do not suffer from any pathologies.
}

\begin{document}
\maketitle
\section{Introduction}

Primordial inflation is one of the most compelling theoretical explanations for creating the seed perturbations for the
large scale structure of the universe and the temperature anisotropy in cosmic microwave background radiation~\cite{WMAP}.
Since inflation dilutes everything, the end of inflation is also responsible for creating the relevant matter required for the Big Bang Nucleosynthesis besides matching the perturbations observed in the CMB, see for instance~\cite{Allahverdi:2006iq,Allahverdi:2011aj}, where inflation has been embedded completely within a supersymmetric Standard Model gauge theory which satisfies all the observed criteria, for a recent review on various models of inflation, see Ref.~\cite{Mazumdar:2010sa}.

In spite of the great successes of inflation, it still requires an ultra-violet (UV) completion. As it stands, within Einstein's gravity inflation cannot alleviate the Big Bang singularity problem, rather it pushes the singularity backwards in time~\cite{Borde,Linde}. In a finite time in the past the particle trajectories in an inflationary space-time abruptly ends, and therefore inflationary trajectory cannot be made past eternal~\cite{Guth}.  Within the framework of  Friedmann--Lema\^{i}tre--Robertson--Walker (FLRW) cosmology, this can be addressed by ensuring the inflationary phase to be preceded either by an ``emergent'' phase where the space-time asymptotes to a Minkowski space-time in the infinite past~\cite{emerge}, or by a non-singular bounce where a prior contraction phase gives way to expansion. Unfortunately, within the  context of General Relativity (GR), as long as the average expansion rate in the past is  greater than zero, i.e. $H_{av}>0$, the standard {\it singularity theorems} due to Hawking and Penrose hold,  and the fluctuations grow as the universe approaches the singularity in the past, which inevitably leads to a collapse of the space-time~\cite{Hawking} commonly referred to as the Big Bang (see also~\cite{Guth}). Thus, in order to provide a geodesic completion to inflation one needs to modify GR.\footnote{For general reviews of modified gravity cosmological models see for example~\cite{Koivisto-rev}.}

There have been some recent progress in ameliorating the UV properties of gravity in the context of a class of ``mildly'' non-local~\footnote{By non-local we only mean that the Lagrangian contains an infinite series of derivatives and hence can only be determined if one knows the function around a finite region surrounding the space-time point in question. This property also manifests in the way one is often able to replace the infinite differential equation of motion with an integral equation~\cite{Zwiebach}. The theories however are only mildly non-local in the sense that the physical variables and observables are local, unlike some other theories of quantum gravity/field theory whose fundamental variables are themselves non-local,  being defined via integrals.} higher derivative theory of gravity~\cite{BMS,BKM,BGKM}, involving terms that are quadratic in the Reimann curvature but contains all orders in higher derivatives. It was found that such an extension of Einstein's  gravity renders gravity {\it ghost-free} and {\it asymptotically-free} in the $4$ dimensional Minkowski space-time.\footnote{For attempts along these lines with a finite set of higher derivative terms, see~\cite{Stelle,Zwiebach85,deser}. However in four dimensions they were not successful in avoiding ghosts and obtaining asymptotic freedom simultaneously, due to essentially the Ostrogradski constraint~\cite{ostrogradski}.} As a result the ordinary massless graviton sees a modified interaction and propagation at higher energies, but at low energies in the infra-red (IR) it exactly yields the properties of Einstein's gravity. It has been demonstrated that such a modification of Einstein's gravity  ameliorates the UV behaviour of the Newtonian potential, and  also permits symmetric non-singular oscillating solutions around the Minkowski space-time~\cite{BGKM} signalling that the action can resolve the cosmological singularities, see also~\cite{BMS,BKM,linearized}.
Similar non-local higher derivative theories of gravity have also been considered in connection with black holes~\cite{warren}, inflationary cosmology~\cite{inflation}, string-gas cosmology~\cite{sgc} and in efforts to understand the quantum nature of gravity~\cite{nlgravity}.\footnote{For investigations along these lines in the more broader class of non-local gravity theories which include non-analytic operators such as $1/\Box$, see~\cite{One_over_Box,oobBI} and references therein.} There was also a proposal to solve the cosmological constant
problem by a non-local modification of gravity~\cite{ArkaniHamed:2002fu} (see also~\cite{Barvinsky2003}).

What is rather intriguing is that non-local actions with an infinite series of higher derivative terms as a series expansion in $\alpha'$, the string tension, have also appeared in stringy literature. One mostly finds them in the open string sector, notably in toy models such as p-adic strings~\cite{padic_st}, zeta-strings~\cite{ZS}, strings on random lattice~\cite{douglas,marc}, and in the open string field theory action for the tachyon~\cite{sft}. For more general aspects of string field theory see for instance~\cite{sft_review}. There have also been progress in understanding quantum properties of such theories, \ie how to obtain loop expansion in the string coupling constant $g_s$, which has lead to some surprising insights into stringy phenomena such as thermal duality~\cite{BJK}, phase transitions~\cite{abe} and Regge behavior~\cite{marc}. It is indeed a wishful thinking to derive such an action in the  {\it closed string}  sector involving gravity.\footnote{We note that significant progress has been made in studying non-local stringy scalar field models coupled minimally and non-minimally to the Einstein gravity \cite{Non-local_scalar}-\cite{KV}.}

In Ref.~\cite{BMS}, with the means of an exact solution, it was shown how a subset of the actions that have been proposed more recently in Ref.~\cite{BGKM}, can ameliorate the Big Bang cosmological singularity of the FLRW metric. The action discussed in~\cite{BMS} only contained the Ricci scalar and it's derivatives up to arbitrary orders and yielded non-singular bouncing solution of {\it hyperbolic cosine } type. However, it was not clear how generic these solutions were, whether these solutions were stable under small perturbations, and whether these theories contain other singular solutions or not.

Some of these issues, in particular concerning only {\it  time dependent} fluctuations around a bouncing solution were addressed in Ref.~\cite{BKM}, where the authors have analysed the stability of the background trajectory for very long wavelength (super-Hubble) perturbations,  either when the  cosmological constant is positive, i.e. $\Lambda >0$, or negative, i.e. $\Lambda <0$.\footnote{Our preliminary investigations indicate that when $\La=0$, the higher derivative modifications allows one to realize the emergent universe scenario.} The latter provided a way to construct cyclic cosmologies, and in particular cyclic inflationary scenarios that have been studied in~\cite{CI1}, and in \cite{CI2} where cosmological imprints were also discussed.

In this paper we aim to focus on the case when $\Lambda >0$. This particular case is very interesting from the point of view of inflation, as it paves the way for a {\it geodesically} complete paradigm  of inflation in past and future. It was found that there exists a de Sitter solution as an attractor for large enough times before and after the bounce. The weakening of gravity in the UV regime facilitates the bounce for the FLRW metric. Our main goal will be to study what happens to the non-singular background solution under arbitrary classical and quantum fluctuations (super and sub-Hubble type) during the bounce  period. We have a two-fold aim. First, we want to verify that at late times there are no fluctuations that are growing during inflation. This would clearly signal an instability in the system which would certainly jeopardize the inflationary mechanism of creating superHubble fluctuations which becomes a constant at late times. Second, we want to check that the bounce mechanism itself is robust, \ie the perturbations do not grow unboundedly near the bounce. In the process of our investigations, we will develop techniques to deal with non-local perturbation equations that one encounters in these theories, and this we believe is a major achievement of the paper. We hope that this will help us in analyzing other cosmological issues in future. Finally, let us comment on a rather interesting feature we noted. It turns out that the requirement of the bounce automatically introduces an extra scalar degree of freedom, and if the mass of this scalar is light, it may be able to play the role of a curvaton \cite{Mazumdar:2010sa}. We however leave a more detailed phenomenological investigation of such a scenario for future.

The paper is organized as follows: In section~\ref{sec:background}, we review some of the results for non-local gravity models and generalize some of the techniques previously used to find exact solutions. In section~\ref{sec:perturbations}, we derive the non-local perturbation equations for the Bardeen potentials around a large class of cosmological background solutions. Next, in section~\ref{sec:late-time} we proceed to solve these equations  around the ``inflationary'' de Sitter late-time attractor solutions that exist in these models. In particular, we obtain the criteria for which the perturbations freeze out or decay after crossing the Hubble radius which allows one to maintain the inflationary mechanism of producing scale-invariant CMB fluctuations. Thereafter, we look at the stability of the fluctuations in section~\ref{sec:bounce} around the bounce. Finally in section~\ref{sec:conclusion}, we conclude by summarizing our results and discussing their implications for inflationary cosmology.


\section{String-inspired Non-local Gravity}\label{sec:background}
In this section we begin by introducing the String theory motivated non-local gravity models. We will then generalize the methods previously employed in~\cite{BMS,BKM} to obtain exact solutions in these theories. We will explore the various non-singular cosmologies that are present in these theories and in particular focus on solutions which can geodesically complete the inflationary space-time. Although  the inflaton potential energy does not remain a strict constant and changes slowly, for technical reasons in our analysis we will treat the potential energy as a ``cosmological'' constant. This should not however affect the issues we plan to address in this paper.
\subsection{Action and Equations of Motion}
In Refs.~\cite{BMS,BKM} the cosmology of a special class of ghost free actions was investigated. The action has the form
\begin{equation}
 S=\int d^4x\sqrt{-g}\left(\frac
 {M_P^2}{2}R+\frac{\lambda}{2}R\Fc(\Box/M_*^2)R-\Lambda+ \mathcal{L}_\mathrm{M}\right),
 \label{nlg_action}
\end{equation}
where   $M_P$ is the Planck mass: $M_P^2=1/(8\pi G_N)$, $G_N$ is the Newtonian gravitational constant,
$\Lambda$  is the cosmological constant, $M_{\ast}$ is the mass scale at which the higher derivative
terms in the action become important, and $\lambda$ is essentially a book keeping device to help us keep track of the higher derivative corrections. $\mathcal{L}_\mathrm{M}$ is the matter
Lagrangian. We use the convention where the metric $g$ has the signature $(-,+,+,+)$.

The analytic function $\Fc(\Box/M_*^2)=\sum\limits_{n\geqslant0}f_n\Box^n$ is an
ingredient inspired by String Theory as an expansion in the string tension ($\al'$). We recall that the action of the covariant d'Alembertian on a scalar is given by
\begin{equation*}
\label{BOX} \Box \equiv g^{\mu\nu}\D_\mu\D_\nu=g^{\mu\nu}\D_\mu\pd_\nu= \frac{1}{\sqrt{-g}} \partial_\mu \left(
\sqrt{-g} \, g^{\mu
  \nu}\partial_\nu \right),
\end{equation*}
where $\D_\mu$ is the covariant derivative. Variation of action Eq.~(\ref{nlg_action}) yields the following equations:
\begin{equation}
\begin{split}
[M_P^2+2\lambda\Fc(\Box)R]G^\mu_\nu
&=\frac{\lambda}{2}\sum_{n=1}
^\infty
f_n\sum_{l=0}^{n-1}\Bigl[g^{\mu\rho}\pd_\rho\Box^l  R  \pd_\nu\Box^{n-l-1}  R
+g^{\mu\rho}\pd_\nu\Box^l  R  \pd_\rho\Box^{n-l-1}  R  {}\\
&{}-\delta^\mu_{\nu}\left(g^{\rho\sigma}
\pd_\rho\Box^l  R  \pd_\sigma\Box^{n-l-1}  R  +\Box^l  R  \Box^{n-l}  R
\right)\Bigr]-\frac{\lambda}{2}
 R \Fc(\Box) R\delta^\mu_{\nu}{}\\&{}+2\lambda(g^{\mu\rho}\D_\rho\pd_\nu-\delta^\mu_{\nu}
\Box)\Fc(\Box) R-\Lambda \delta^\mu_{\nu}+{T_{\mathrm{M}}}^\mu_\nu ,
\end{split}
\label{eqEinsteinRonlyupdown}
\end{equation}
where ${T_{\mathrm{M}}}^\mu_\nu$ is the energy--momentum tensor of
matter and
\begin{equation}
G^\mu_\nu=R^\mu_\nu-\frac12\delta^\mu_\nu R\ ,
\end{equation}
is the Einstein tensor. We assume that the matter stress energy tensor obeys the usual conservation equation, $\D_\mu{T_{\mathrm{M}}}^\mu_\nu=0$, to ensure consistency of the modified Einstein's equations, which we will refer to in future as Einstein--Schmidt (ES) equations~\footnote{The generalization of the Einstein's equations to include terms such as $R\Box^nR$ was analyzed by Schmidt~\cite{Schmidt}, having an infinite series of such terms give rise to Eq.~(\ref{nlg_action}).}. The ES equations can be written in a reasonably compact form:
\begin{equation}
\begin{split}
&[M_P^2+2\lambda\Fc(\Box)R]G^\mu_\nu={T_{\mathrm{M}}}^\mu_\nu-\Lambda \delta^\mu_{\nu}+{}\\
+&\lambda\Kc^\mu_\nu -\frac{\lambda}{2}\delta^\mu_{\nu}\left(\Kc^\sigma_\sigma+\Kc_{1}
\right)-\frac{\lambda}{2}
 R \Fc(\Box) R\delta^\mu_{\nu}+2\lambda(g^{\mu\rho}\D_\rho\pd_\nu-\delta^\mu_{\nu}
\Box)\Fc(\Box) R \ ,
\end{split}
\label{eqEinsteinRonlyupdownshort}
\end{equation}
where we have introduced two additional quantities:
\begin{equation}
\Kc^\mu_\nu=g^{\mu\rho}\sum_{n=1}
^\infty
f_n\sum_{l=0}^{n-1}\pd_\rho\Box^l  R  \pd_\nu\Box^{n-l-1}  R\,,\quad\quad\quad
\Kc_1=\sum_{n=1}
^\infty
f_n\sum_{l=0}^{n-1}\Box^l  R  \Box^{n-l}  R\,.
\label{KKK}
\end{equation}
Making any progress with the ES equations may seem daunting at first, but previous studies have shown that the trace equation can be tractable since all we have to deal with is the Ricci scalar:
\begin{equation}
M_P^2R- \lambda\sum_{n=1}^\infty
f_n\sum_{l=0}^{n-1}\Bigl(\pd_\mu\Box^l  R  \pd^\mu\Box^{n-l-1}  R
+2\Box^l  R  \Box^{n-l}  R\Bigr)-  6\lambda\Box\Fc(\Box)R=4\Lambda-{T_{\mathrm{M}}}^\mu_\mu\, ,
\label{eqEinsteinRonlytrace}
\end{equation}
or, in the more compact notation
\begin{equation}
M_P^2R-\lambda\Kc^\mu_\mu-2\lambda\Kc_1- 6\lambda\Box\Fc(\Box)R=4\Lambda-{T_{\mathrm{M}}}^\mu_\mu\, ,
\label{eqEinsteinRonlyupdownNctrace}
\end{equation}


\subsection{General Ansatz for finding Exact Solutions}
It has been shown in Refs.~\cite{BMS,BKM,KV_BounceSOl} that the following ansatz
\begin{equation}
\Box R-r_1R-r_2=0\mand r_1\neq 0 \ ,
 \label{ansatz2}
\end{equation}
is useful in finding exact solutions. From Eq.~(\ref{ansatz2}) one gets the following
relationships:
\begin{equation}
\begin{split}
\Box^n R&=r_1^n\left(R+\frac{r_2}{r_1}\right)\quad\text{for}\quad n>0,\\
\Fc(\Box)
R&=\Fc_1R+\frac{r_2}{r_1}(\Fc_1-f_0)\quad\text{where}\quad \Fc_1\equiv\Fc(r_1).
\end{split}
 \label{ansatz_cons}
\end{equation}
If the scalar curvature $R$ satisfies (\ref{ansatz2}), then equations (\ref{eqEinsteinRonlyupdown}) have the following form:
\begin{equation}
\begin{split}
&\left[M_P^2+2\lambda\left(\Fc(r_1)R+\frac{r_2}{r_1}(\Fc(r_1)-f_0)\right)\right]
G^\mu_\nu={T_{\mathrm{M}}}^\mu_\nu\\
+&\lambda\Fc^{(1)}(r_1)\left[\pd^\mu R \pd_\nu
R-\frac{\delta^\mu_\nu}{2}\left(g^{\sigma\rho}\pd_\sigma R \pd_\rho
R+r_1\left(R+\frac{r_2}{r_1}\right)^2\right)\right]-\Lambda \delta^\mu_\nu
{}\\+
&2\lambda\Fc_1\left[D^\mu\pd_\nu
R-\delta^\mu_\nu(r_1R+r_2)\right]-\lambda\frac{\delta^\mu_\nu}{2}\left[\Fc_1R^2-\frac{r_2^2}{r_1^2}(\Fc_1-f_0)
\right]
\end{split}
\label{eqEinsteinAnzatz}
\end{equation}
where $\Fc^{(1)}$ is the first derivative  of $\Fc$ with respect to the argument, and
the trace equation (\ref{eqEinsteinRonlytrace}) becomes especially simple:
\begin{equation}
A_1R-\lambda\Fc^{(1)}(r_1)\left(2r_1R^2+ \partial_\mu R \partial^\mu R\right)+A_2=-{T_{\mathrm{M}}}^\mu_\mu\,,
\label{eqEinsteinRonlytraceansatz2}
\end{equation}
where
\begin{equation*}
\begin{split}
A_1&={}M_P^2-\lambda\left(4\Fc^{(1)}(r_1)r_2-2\frac{r_2}{r_1}(\Fc_1-f_0)+6\Fc_1r_1\right) \mand\\
A_2&={}-4\Lambda-\lambda\frac{r_2}{r_1}\left(2\Fc^{(1)}(r_1)r_2-2\frac{r_2}{r_1}(\Fc_1-f_0)+6\Fc_1r_1\right).
\end{split}
\end{equation*}
We proceed to consider a traceless radiation along with a cosmological constant. This will simplify many of our calculations.
Then the simplest way to obtain a solution to Eq.~(\ref{eqEinsteinRonlytraceansatz2}) with ${T_{\mathrm{M}}}^\mu_\mu=0$ is
to put $A_1=A_2=0$, and impose 
\be
\Fc^{(1)}{(r_1)}=0\ .
\label{r1}
\ee
This implies
\begin{equation}
 r_2={}-\frac{r_1[M_P^2-6\lambda\Fc_1r_1]}{2\lambda[\Fc_1-f_0]},
\qquad\qquad\qquad\qquad
\Lambda={}-\frac{r_2M_P^2}{4r_1}. \label{r2lambda}
\end{equation}
Thus we see that within the context of the ansatz,  the ES equations and the corresponding solutions are characterized by three independent parameters, $r_1$, $ \lambda f_0$, and $ \lambda \cF_1$. Although $\la$ is only a book-keeping device, and one can set it to one if desired,  it is interesting to look at the analytic properties of the different parameters  with respect to $\la$ since the  $\la\ra 0$ limit can mimic the large time limit when one expects to recover GR and all the higher derivatives become  small. Now,  $r_1$ is determined by (\ref{r1}) and  is therefore independent of $\la$, but $r_2$  depends non-analytically on $\la$.\footnote{Actually, not only $r_2$, $\La$ also acquires a non-analytic dependence through the second equation in (\ref{r2lambda}). In fact, this has to be the case, otherwise we could take the $\la\ra0$ limit smoothly and conclude that GR has non-singular bouncing solutions!} This could seem to be a potential problem in recovering General Relativistic solutions. As we will see later in explicit examples, although $r_2$ may have non-analytic dependence, it is really just an intermediate variable of convenience, and the solutions for the scale-factor need not have any singular dependence in $\la$ (see section~\ref{sec:hyperbolic}). This is why one is able to recover the GR solutions in appropriate limits.

It is important to note that by virtue of the Bianchi identities, solving the trace of the ES equations
essentially guarantee that we have a solution for the complete ES
equations, which now  simplifies to
\begin{equation}
\begin{split}
&2\lambda\Fc_1(R+3r_1)
G^\mu_{\nu}
={T_{\mathrm{M}}}^\mu_\nu+
2\lambda\Fc_1\left[g^{\mu\rho}\D_\rho\pd_\nu
R-\frac14\delta^\mu_\nu\left(
R^2+4r_1R+r_2\right)\right] \ .
\end{split}
\label{eqEinsteinAnzatz_Ai0}
\end{equation}
In general one
is required to include radiative sources to get exact solutions~\cite{BKM}, and we need to make
sure that the radiative energy density if present is positive and not ghost-like.
We will discuss this point in the specific context of cosmological
solutions in the following subsection.

Let us emphasize that equations up to this point are general and do not take into account the
properties of the metric. Thus the above procedure can be used to obtain  metrics which are cosmological  (time dependent solutions) in nature or which have spatial dependence. This is an important generalization on the analysis discussed in~\cite{BKM}, and a major step forward in investigating these non-local models.

\subsection{Cosmological Solutions, Bounce and Stability}\label{sec2:stability}

We are now going to focus on spatially flat FLRW type cosmological solutions that was obtained in~\cite{BMS,BKM} in the presence of a cosmological constant and radiation energy density,
\begin{equation}
\rho=\rho_0\left(\frac{a_0}{a}\right)^4
\label{matter13}
\end{equation}
with a traceless energy momentum tensor. In the above expression $a$ is the scale factor and $a_0$ refers to its value at the bounce. In the cosmological context  the ansatz Eq.~(\ref{ansatz2}) becomes a third order differential equation for the Hubble parameter, $H=\dot a/a$,
\begin{equation}
\label{EquH}
 \dddot H+7H\ddot H+4\dot H^2+12H^2\dot H
 ={}-2r_1H^2-r_1\dot H-\frac{r_2}{6}\ ,
\end{equation}
where dot defines derivative w.r.t physical time. The solutions to the above equation, their stability, and asymptotic behaviours have been studied in details in~\cite{BKM}. It was found that several non-singular solutions exist where the universe bounces from a phase of contraction to a phase of expansion, thus providing a resolution to the Big Bang singularity problem. By looking at the $'00'$ component of the ES equations, one can calculate the radiation energy density at the bounce point, $\rho_0$, which must be positive:
\begin{equation}
\rho_0=\frac{3(M_p^2r_1-2\la f_0 r_2)(r_2-12h_1 M_*^4)}{12r_1^2-4r_2},
\label{FRW_eqEinsteinRonlyansatztt}
\end{equation}
where $h_1=\ddot{H}/M_*^3$ characterizes the acceleration of the universe at the bounce point and plays the role of an `initial condition'. There are three cosmological scenarios one can envisage from Eq.~(\ref{EquH}) :

\begin{itemize}
\item $\La<0,\ r_1>0\Ra r_2>0$: One obtains a cyclic universe scenario where the universe undergoes successive bounces and turnarounds. Qualitatively speaking the bounce is caused by the higher derivative modifications, while the turnarounds occur when the radiation energy density cancels the negative cosmological constant. Such scenarios can give rise to ``cyclic inflationary'' models if one is able to incorporate entropy production~\cite{CI1,CI2}, but we will not consider these scenarios here any further.

\item $\La>0,\ r_1<0\Ra r_2>0$: One obtains a bouncing universe scenario where a phase of contraction gives way to a phase of super-inflating expansion phase. An exact example is the Ruzmaikin-type solution~\cite{Ruzmaikin,Toporensky},
which incidentally contains no radiation~\cite{KV_BounceSOl}. Unfortunately, in these scenarios the universe is stuck in a quantum gravitational phase accelerating faster and faster, and we never recover a GR phase. Clearly such solutions are not phenomenologically viable and hence we disregard these scenarios.

\item $\La>0,\ r_1>0\Ra r_2<0$: This represents a geodesic
completion of an inflationary space-time via a non-singular bounce. It is easy to see that the action Eq.~(\ref{nlg_action}) without  matter admits constant curvature vacuum solutions, the de
Sitter solution or the Minkowski solution, depending upon the value of
the cosmological constant. These background solutions are in fact
exactly the same as that of the general relativistic de Sitter phase, as for
constant $R$, all the higher derivative terms in the field equations
vanish, and one just obtains from Eq.~(\ref{eqEinsteinRonlytrace})
\begin{equation}
\label{de SitterSol}
R_{dS}={4\frac{\Lambda}{M_P^2}}.
\end{equation}
This is precisely the de Sitter solution the space-time asymptotes to in the more general case, as the radiation dilutes away very quickly during the cosmological constant dominated phase.
\end{itemize}

It is this last case that is relevant for inflationary cosmology, and is the main focus of our paper. The higher derivative extension of gravity provides us with a plausible answer to the question, ``what happened before inflation?''. The general picture that emerges is that in the presence of a positive cosmological constant, if a small causal patch is approximately homogeneous and isotropic at Planckian densities, the universe would undergo a non-singular bounce, following which the universe will start to inflate quickly diluting any matter (radiation) present in the system, and we essentially will have a pure de Sitter. Of course, to be consistent with the inflationary phenomenology, $\La$ cannot be a strict constant but has to evolve, possibly with a scalar field, the {\it inflaton}~\cite{Mazumdar:2010sa}. For technical reasons we haven't been able to incorporate this dynamics yet, but our main aim here is to assess the robustness of the bounce mechanism and the inflationary scenario in terms of perturbations.

What we expect is that with the dilution of radiation energy density during the vacuum energy dominated phase, any fluctuations present in the fluid would also get diluted away, so that the fluctuations can become dominated by the quantum fluctuations of the inflaton field which we know can produce the observed approximately scale-invariant spectrum in CMB. Since we do not have any inflaton in our scenario, what we want to check  is whether  there are any growing or unstable modes of fluctuations around the geodesically complete de Sitter space-time.  If present, this would signal either the fallibility of the bounce mechanism, or  the breakdown of the inflationary mechanism of generating fluctuations.

\subsection{The Hyperbolic Cosine Bounce}
\label{sec:hyperbolic}
It is rather fortunate that one can actually find an exact solution for the higher derivative extension of gravity under consideration, which serves as a good illustration of a geodesic complete space-time.
It is given by
\begin{equation}
a(t) = a_0\cosh{\sqrt{\frac{r_1}{2}}t},\quad\Rightarrow \quad H=\sqrt{\frac{r_1}{2}}\tanh\left(\sqrt{\frac{r_1}{2}}t\right)
\label{FRW_Hexact}
\end{equation}
 This solution satisfies the ansatz with the specific parameter combination
\begin{equation*}
\Box R = r_1 R-6r_1^2.
\end{equation*}
Substituting $r_2=-6 r_1^2$ into Eqs.~(\ref{r2lambda}) and (\ref{FRW_eqEinsteinRonlyansatztt}) we get
\begin{equation}
\label{lambdaMp}
\Lambda=\frac{3}{2}r_1M_P^2 \quad\mand\quad \rho_0=-\frac{27}{2}\la\Fc_1r_1^2.
\end{equation}
Note that the radiation is non-ghost like, provided
\begin{equation}
\Fc(r_1)<0.
\label{eqEinsteinRonlytraceansatzAA20_exactF}
\end{equation}
This turns out to be a general constraint which must be satisfied in order to realize geodesically complete de Sitter space-time, and we are going to assume that this constraint is satisfied. The above specific example (\ref{FRW_Hexact}) serves as a nice illustration of some of our general results and we will fall back on it as appropriate in the remainder of the paper.

Finally, one may wonder about the special case when $\Fc_1=0$, which means there
is no radiation. Although this is under current investigations, indications are that such theories will suffer from the presence of ghosts. Effectively there is a double pole in the propagator at $p^2=r_1$.

\section{Perturbations}\label{sec:perturbations}

In the previous section we have provided a general algorithm for finding cosmological background solutions in non-local gravity models which in particular can realize inflationary space time at late times. We have also provided concrete examples of such solutions. What we now want to address is the robustness of the paradigm by studying the fluctuations around the background. This will let us test whether the bounce mechanism is stable, as well as whether the inflationary mechanism of generating scale-invariant perturbations can survive in the presence of higher derivative modifications. In this section we will provide a general framework for dealing with the scalar modes and obtain the coupled (non-local) differential equations describing the evolution of the Bardeen potentials. Since this is a somewhat technical section, we have provided a summary subsection~\ref{summary}, where we enumerate the results in a self-contained manner.

\subsection{Perturbations on higher derivative action of gravity}

To compute perturbation equations we use the fact that the background configuration obeys the ansatz (\ref{ansatz2}) and conditions (\ref{r2lambda}). Let us introduce the following notations to separate the background and the perturbations:
\begin{equation}
 R=R_B+\delta\! R,\qquad~~~~~ \Box={\Box}_B+\delta\Box,
\end{equation}
where the subscript $B$ stands for background. Also, let us denote the variation of the ansatz (\ref{ansatz2})  as follows:
\begin{equation}\label{zeta}
    \zeta\equiv\delta\Box R_B+(\Box_B-r_1)\delta\!R.
\end{equation}
While computing the variation of the ES equations, what we encounter most often is the variation $\delta(\Box^n R)$,
which thanks to the ansatz (\ref{ansatz2}) sums up to
\begin{equation}
\delta(\Box^n R)=\frac{\Box_B^n-r_1^n}{\Box_B-r_1}\delta\Box R_B+\Box_B^n\delta R=\frac{\Box_B^n-r_1^n}{\Box_B-r_1}\zeta+r_1^n\delta\!R\,,
\label{deltaRn}
\end{equation}
and this will turn out to be the key simplification~\footnote{Similar relationships are useful for the perturbation analysis in models with non-local scalar field as shown in Ref.~\cite{KV}.}.
This can be used to obtain
\begin{equation}
\delta(\Fc(\Box) R)=\frac{\Fc(\Box_B)-\Fc_1}{\Box_B-r_1}\zeta+\Fc_1\delta\!R=(\Box_B-r_1)\Xi+\Fc_1\delta\!R\,,\label{deltaFR}
\end{equation}
where we have introduced
\begin{equation*}
\Zc(\Box)=\frac{\Fc(\Box)-\Fc_1}{(\Box-r_1)^2},\qquad~~~~~~~
\Xi=\Zc(\Box_B)\zeta.
\end{equation*}
Note that, expanding $\Fc(\Box)$ in the Taylor series, one can see that $\delta(\Fc(\Box) R)$ has no pole. Also thanks to
 $\Fc^{(1)}(r_1)=0$,  $\Zc(\Box)$ is an analytic function. Similarly, we obtain
\begin{eqnarray}
\label{deltaBoxFsR}
    \delta(\Box\Fc(\Box)R) &=& \Box_B(\Box_B-r_1)\Xi+\Fc_1(\zeta+r_1\delta\!R),\\
\delta\Kc^\mu_\nu&=&\left(\pd^\mu\Xi\pd_\nu R_B+\pd^\mu R_B\pd_\nu\Xi\right), \\
\delta\Kc_1&=&2\Xi\left(r_1R_B+r_2\right)+R_B(\Box_B-r_1)\Xi-\frac{r_2}{r_1}\delta R(\Fc_1-f_0)\ ,
\label{deltasKKKFprime0}
\end{eqnarray}
where we have used conditions Eq.~(\ref{r2lambda}), and in particular $\Fc'(r_1)=0$, to obtain Eq.~(\ref{deltasKKKFprime0}).

\subsection{Perturbing the trace equation }
Assuming $\delta {T_\text{M}}^\mu_\mu=0$, which is true for radiation, we get the trace equation (\ref{eqEinsteinRonlyupdownNctrace})
for perturbations linearized about a background solution:
\begin{equation}
M_P^2\delta\!R-\lambda\delta\Kc^\mu_\mu-2\lambda\delta\Kc_1- 6\lambda\delta(\Box\Fc(\Box)R)=0.
\label{eqEinsteinRonlyupdownNctracedelta}
\end{equation}
Substituting Eqs.~(\ref{deltaBoxFsR}) and (\ref{deltasKKKFprime0}), and using condition $A_1=0$ to eliminate $M_P^2$,
we obtain
\begin{equation}
\pd^\mu\Xi\pd_\mu R_B+r_1R_B\Xi+2r_2\Xi+R_B\Box_B\Xi+
3[\Box_B(\Box_B-r_1)\Xi+\Fc_1\zeta]=0.
\label{deltatraceGI}
\end{equation}
This is a very important equation because it gives us information about $\zeta$.
It can be rewritten as
\begin{equation}
\Pc\zeta=0,
\label{deltatraceGInonconst}
\end{equation}
where
\ba
\Pc&=&[\pd^\mu R_B\pd_\mu+2(r_1 R_B+r_2)]\Zc(\Box_B)+\Wc(\Box_B)
\ea
with
\begin{equation*}
\Wc(\Box_B)=3\Fc(\Box_B)+(R_B+3r_1)\frac{\Fc(\Box_B)-\Fc_1}{\Box_B-r_1}\,.
\label{deltatraceGIconst}
\end{equation*}
As we see, operator $\Pc$ depends not only on d'Alembertian but also on partial derivatives and has non-constant coefficients.
Nevertheless, we have been able to find a non-local differential equation for one of the perturbative variable $\za$, which one can use different techniques to solve. The form of the perturbed trace equation demonstrates the importance of the function $\zeta$.
\subsection{Scalar Perturbations}
The metric perturbations can be divided into $4$ scalar, $4$ vector and $2$ tensor degrees of freedom, according to their transformation
properties with respect to the three spatial coordinates on
the constant-time hypersurface. Different types of perturbations do
not mix at the first order~\cite{Lifshitz1946}-\cite{hwangnoh}. In this paper, we are going to focus on the scalar perturbations leaving the detailed analysis of vector and tensor modes for separate projects. Scalar metric perturbations are given by $4$ arbitrary scalar
functions $\phi(\tau,x^a)$,~$\beta(\tau,x^a)$,~ $\psi(\tau,x^a)$,~ $\gamma(\tau,x^a)$ in the following way
\begin{equation}
ds^2=a(\tau)^2\left[-(1+2\phi)d\tau^2-2\pd_i \beta d\tau
dx^i+(\{1-2\psi \}\delta_{ij}+2\pd_i\pd_j\gamma)dx^idx^j\right],
\label{mFr}
\end{equation}
where $\tau$ is the conformal time related to the cosmic one as
$a(\tau)d\tau=dt$.
To avoid gauge fixing issues we use gauge-invariant variables. There exist two
independent gauge-invariant variables (the Bardeen potentials), which
fully determine the scalar perturbations of the metric
tensor~\cite{Bardeen}-\cite{hwangnoh}
\begin{equation}
\Phi=\phi-\frac{1}{a}(a\vartheta)^\prime=\phi-\dot \chi,\qquad\Psi=\psi+\Hc\vartheta=\psi+H\chi,
\label{GIvars}
\end{equation}
where  $\chi=a\beta+a^2\dot\gamma$, $\vartheta=\beta+\gamma'$, $\Hc(\tau)=a'/a$, and differentiation with respect to the
conformal time $\tau$ is denoted by a prime.

Explicit calculations show that $\zeta$ is a gauge-invariant function, and it is therefore not possible to choose such gauge that  $\zeta=0$. Now, the function $\zeta$ can be written as
\begin{equation}
\zeta=\frac{R_B'}{a^2}(\Phi'+3\Psi')-2(r_1 R_B+r_2)\Phi +\left(\Box_B-r_1\right)\delta\!R_{\text{GI}}.
\label{deltaFRGI}
\end{equation}
where $\delta\!R_{\text{GI}}$ is the gauge invariant part of the curvature variation:
\begin{equation}
\delta\!R_{\text{GI}}= \frac{2}{a^2}\left[k^2 (\Phi-2\Psi)-3\frac{a'}{a}\Phi'-6\frac{a''}{a}\Phi-3\Psi''-9\frac{a'}{a}\Psi'\right].
\label{deltaRGI}
\end{equation}
and we have used
\begin{equation}
\delta\Box=\frac{1}{a^2}\left[2\Phi\left(\pd_\tau^2+2\frac{a'}{a}\pd_\tau\right)+\left(\Phi'+3\Psi'\right)\pd_\tau\right].
\end{equation}
Hereafter $k$ is the comoving wavenumber.
The formula (\ref{deltatraceGInonconst}) thus provides us with one equation for $\Phi$ and $\Psi$.

It is essential to get another relationship among the Bardeen potentials and one can accomplish this task by considering the $(i\neq j)$ components of the perturbed ES equations, Eq.~(\ref{eqEinsteinRonlyupdownshort}). Trivially, all the terms with $\delta^i_j$ go away, $\delta\Kc^i_j=0$ and $\delta\Gamma^0_{mn}=0$, and only few non-trivial pieces survive.
After some algebra  we get the following equation
\begin{equation}
(\Box_B-r_1)\Zc(\Box_B)\zeta+\Fc_1[\delta R_{\text{GI}}+(R_B+3r_1)(\Phi-\Psi)]=0.
\label{deltaGI1lambdanopis}
\end{equation}
This provides us with the second equation which we need in order to solve for the Bardeen potentials.

It is curious to note that for radiative sources we were able to decouple the evolution equations for the
the metric perturbations from  the fluctuations in fluid. For the sake of completeness we have provided the equations that determine the fluid fluctuations from the Bardeen potentials in the appendix.
\subsection{Perturbation summary}\label{summary}
In this section we are able to summarize two differential equations for the two Bardeen potentials, $\Phi$ and $\Psi$. They read as follows:
\ba
[\pd^\mu R_B\pd_\mu+2(r_1 R_B+r_2)]\Zc(\Box)\zeta+\Wc(\Box)\zeta&=&0\label{eq341}\\
(\Box_B-r_1)\Zc(\Box)\zeta+\Fc_1[\delta R_\text{GI}+(R_B+3r_1)(\Phi-\Psi)]&=&0
\label{eq342}
\ea
where
\ba
\zeta&=&\delta\Box R_B+(\Box_B-r_1)\delta\!R_\text{GI}\nonumber\\
\Box_B &=&-\frac{1}{a^2}\pd_\tau^2-2\frac{a'}{a^3}\pd_\tau-\frac{k^2}{a^2}=-\pd_t^2-3\frac{\dot a}{a}\pd_t-\frac{k^2}{a^2}\nonumber\\
\delta\Box&=&\frac{2}{a^2}\Phi\pd_\tau^2+\frac{4a'}{a^3}\Phi\pd_\tau+\frac{1}{a^2}(\Phi'+3\Psi')\pd_\tau\nonumber\\
\delta\!R_\text{GI}&=& \frac{2}{a^2}\left(k^2 (\Phi-2\Psi)-3\frac{a'}{a}\Phi'-6\frac{a''}{a}\Phi-3\Psi''-9\frac{a'}{a}\Psi'\right)\nonumber\\
&=&6\Box_B\Psi-2R_B\Phi-6\frac{a'}{a^3}(\Phi'+\Psi')+2\frac{k^2}{a^2}(\Phi+\Psi)\label{delta-rg}\nonumber\\
\Wc(\Box)&=&3\Fc(\Box)+(R_B+3r_1)\frac{\Fc(\Box)-\Fc_1}{\Box-r_1}\nonumber\\
\Zc(\Box)&=&\frac{\Fc(\Box)-\Fc_1}{(\Box-r_1)^2}\nonumber
\ea
Clearly, equations (\ref{eq341}) and (\ref{eq342}) provide us with a coupled system of equations, albeit non-local, for the two Bardeen potentials, $\Phi,~\Psi$. In our analysis, we never really need the quantities describing the matter fluctuations, but they can indeed be obtained by looking at the other components of the ES equations. We provide some details of that in the appendix.

\section{Perturbations at late-times in the de Sitter limit}\label{sec:late-time}

Evidently it is a rather difficult task to solve the evolution equations for the Bardeen potentials for the most general case, even if  $\zeta$ vanish~\footnote{There have been steady progress in trying to  solve and understand some of the properties of such infinite differential equations, see for instance \cite{math}.}. We can gain however significant insight into how the perturbations evolve within certain approximations. In particular, we are interested in addressing two important questions: (i) do we recover the GR limit during the inflationary phase, this would ensure that the inflationary mechanism of generating near scale-invariant perturbations can be trusted despite the higher derivative modifications to gravity, and  (ii) we also want to see whether any pathologies develop around the bounce, this would lead to robustness of the bouncing mechanism, which several other non-singular alternatives seem to lack.
In this section we will address the first question.

\subsection{Localizing equations}

At later times, $t\ra+\infty$, solution~(\ref{FRW_Hexact}) tends to a de Sitter space-time, with $R\ra 4\La$. This simplifies the perturbation equations (\ref{deltatraceGInonconst}) and (\ref{deltaGI1lambdanopis}) remarkably.
At large times $R\ra -r_2/r_1$, and therefore we get
\begin{equation*}
\zeta \approx  (\Box-r_1)\da R_\text{GI},\quad\Pc(\Box)\approx\Wc(\Box)|_{\text{dS}}
\end{equation*}
where the subscript 'dS' indicates that we substitute $R_B=-r_2/r_1$. Resulting perturbation equations are as follows:
\begin{eqnarray}
\Wc(\Box)|_\text{dS}(\Box-r_1)\da R_\text{GI}&=&0,\label{sec4case11}\\
\Fc(\Box)\delta R_\text{GI}+(-r_2/r_1+3r_1)(\Phi-\Psi)&=& 0.
\label{sec4case22}
\end{eqnarray}
 In this limit $\Wc$ is an analytic function with constant coefficients. We can therefore solve the first equation assuming that we know the roots of the algebraic (or transcendental) equation $\Wc(\omega^2)|_\text{dS}(\omega^2-r_1)=0$~\cite{neil}. In the case of simple roots $\omega_i$ this reduces to\footnote{If some roots  $\omega_i$ are multiple roots they produce equations of the form $(\Box-\omega_i^2)^{m_i}\delta R_\text{GI}=0$, where $m_i$ is the multiplicity, and these must be treated differently. For the case of double roots see for instance \cite{v2}. For simplicity we assume that only simple roots are present.}
\ba
\delta R_{\text{GI}}=0~~~\text{or}~~~(\Box-\omega_i^2)\delta R_\text{GI}=0\label{dscases}
\ea
There is always one root $\omega_1^2=r_1$ but there could be other roots of $\Wc(\omega^2)$,  $\oa_i^2$ with $i=2..N$. Once $\delta  R_\text{GI}$ is found to be, say $\delta R_i$ corresponding to the root $\oa_i^2$, we can express $\da R$ in terms of $\Phi$ and $\Psi$ and solve the equation
\ba
\delta R_{\text{GI}}=\delta R_i
\ea
along with (\ref{sec4case22}). The most general solution is obviously an arbitrary superposition of all the solutions.

Before proceeding further a couple of remarks are in order. Firstly, the fact that solutions to non-local equations  can be approximated as a linear superposition of solutions to a collection of local equations is a rather interesting feature of these models, but it is not surprising and has been discussed in the literature. Secondly, the fact that there always exists a root $\oa_1^2=r_1$ is also not surprising. A straight forward application of the results obtained in~\cite{BGKM} tells us that for actions of the form (\ref{nlg_action}), there is always a scalar degree of freedom, the famous Brans-Dicke theory belongs to this class as well, $r_1$ is simply the mass squared of this scalar. Finally, we note that the roots, their number and values depend on $\cF$, and in particular one can choose $\cF$'s such that there are no roots other than $r_1$.
\subsection{Trivial case: $\delta R_{\text{GI}}=0$}

Trivial possibility of Eq.~(\ref{dscases}) implies
\ba
\Phi-\Psi=0
\label{GR-relation}
\ea
as it is in GR. What we then have to solve is
\ba
\delta R_0=2\left(3\Box_B\Phi-R_B\Phi-6\frac{a'}{a^3}\Phi'+2\frac{k^2}{a^2}\Phi\right)=0
\ea
Substituting the appropriate background functions for de Sitter, and $R_B\equiv 12H^2$, we can solve the above equation in terms of elementary functions
\be
\Phi=\eta[c_1(\cos(\eta)+\eta\sin(\eta))+
c_2(-\eta\cos(\eta)+\sin(\eta))]\ ,
\ee
where
\be
\eta=\frac{k\tau}{\sqrt{3}}=\frac{k}{\sqrt{3}aH}\,.
\ee
and we remind the readers that the inflationary scale factor in conformal time is given by
\be
a(\tau)=-\frac{1}{H\tau}\with \tau\in(-\infty,0)
\ee
At large times clearly $\eta\ra 0$ as the scale factor is increasing exponentially, and all the terms die out. This is precisely the mode that one would obtain in ordinary GR in the presence of a cosmological constant and radiative sources. Within a few e-folds, the radiation density and it's fluctuations become irrelevant and the quantum vacuum fluctuations of the inflaton becomes the dominant source of metric perturbations.
\subsection{Non-trivial case, $\delta R_{\text{GI}}\neq 0$}
We now consider solutions to the equation
\be
(\Box_B-\omega_i^2)\delta R_i=0\,.
\ee
Acting $\Box_B-\omega_i^2$ on  (\ref{sec4case22}) we get
\be
(\Box_B-\omega_i^2)(\Phi-\Psi)=0
\ee
Exact solutions to the above equation are known in terms of the
conformal time:
\be
\Phi_{k}-\Psi_k=
H(-\tau)^{\frac32}[d_{1k}J_{\nu}^1(-k\tau)+d_{2k}Y_{\nu}^2(-k\tau)]
\ee
where $J_{\nu},Y_{\nu}$ are Bessel functions of the first and the second kind
respectively, and
\be
\nu^2=\LF{\frac94}-\frac{\oa_i^2}{H^2}\RF
\ee
Notice that  when all $d_{1k},d_{2k}$'s vanish we recover the well known GR limit $\Phi-\Psi=0$.

We are primarily interested in the late time super-Hubble limit, $k|\tau|\ll 1$, of these modes. The asymptotic forms of the Bessel functions only depend on the real part of $\nu$, lets call it $\nu_R$. Then we have
\ba
&\lim_{x\ra 0}&J_{\nu}(x)\sim x^{\nu_R}\mand\\
&\lim_{x\ra 0}&Y_{\nu}(x)\sim \left\{
\begin{array}{cc}
x^{-\nu_R}&\for \nu_R>0\\
\ln(x) &\for \nu_R=0
\end{array}
\Rd
\ea
We are now in a position to identify the different cases:
\begin{itemize}
\item If $\oa_i^2<0$, then $\nu^2$ is positive, and $\nu>3/2$. This means the $Y_{\nu}$ function grows faster that $(-\tau)^{-3/2}$ as $\tau\ra 0$, which in turn implies that $\Phi-\Psi$ grows at late times making the inflationary trajectory unstable, and the perturbations  unviable for inflationary cosmology.
\item If $\oa_i^2\approx 0$. This is a rather interesting and special case corresponding to the presence of a massless degree of freedom. As it is clear from the asymptotics, in this case $\Phi-\Psi$ tends to a constant at late times. In fact, if one assumes the usual Bunch-Davis type subHubble vacuum initial conditions for this field, then one will end up with a scale invariant spectrum for these modes. This is very similar to the curvaton scenario~\cite{Mazumdar:2010sa} that is routinely considered in inflationary cosmology and we plan to study the phenomenological implications of this mode in more details in the future.
\item If $9H^2/4>\oa_i^2>0$, then $\nu^2$ is positive and $0<\nu<3/2$. Firstly, this means that the $J_{\nu}$ function decays at large times. The $Y_{\nu}$ function does grow but at a rate slower than $(-\tau)^{-3/2}$, so that $\Phi-\Psi$ decays as $\tau\ra0$ leading to the GR relation (\ref{GR-relation}). Inflationary cosmology is thus safe from these modes.
\item If $9H^2/4<\oa_i^2$, then $\nu^2<0$, and $\nu$ is imaginary, \ie $\nu_R=0$. The leading behavior of the two modes in $\Phi-\Psi$ are then given by $(-\tau)^{-3/2}$ and $(-\tau)^{-3/2}\ln(-\tau)$ and they both vanish as $\tau\ra 0^-$.
\item If $\oa_i^2$ is complex, then $\nu^2$ is also complex and whether these modes are dangerous or not depends on the value of $\nu_R$. According to the discussion above, for phenomenological consistency of the model we must have $\nu_R\leq 3/2$.
\end{itemize}

Thus to summarize, our criteria for preserving the inflationary mechanism is that all the roots, $\oa_i$, should be such that the corresponding
\be
|\nu_R|<3/2\ .
\label{criteria}
\ee  For real roots, this simply means that they must also be non-negative. In particular since $\oa^2=r_1$ is always a root, we must have $r_1>0$. As discussed in section~\ref{sec2:stability}, this is consistent with the observations in \cite{BKM},
where it was found that when $r_1<0$, the de Sitter solution is no longer a stable solution, but it is when $r_1>0$.
Therefore, there exists a large class of models which are compatible with inflationary predictions.
\section{Perturbations across the bounce}\label{sec:bounce}
Having determined the late time behavior of the different perturbation modes around our inflationary bouncing solution, our next task is to see how they behave near the bounce. Since we have already provided the criteria to ensure that none of the fluctuations grow during inflation, all we now have to check is that there are no modes which become singular at any finite time. In particular we will focus near the bounce, as within a few e-folds of expansion the scale-factor is expected to resemble the de Sitter phase which we have already analyzed, at least this is what happens in the hyperbolic cosine bounce (\ref{FRW_Hexact}). In fact, for most part we will restrict to this special case to keep the analysis tractable.

\subsection{Bounce Limit}
Near the bounce, even though the equations can again be simplified considerably it is a more complicated limit. The main difference with the previous section is that we can approximate $R_B$ by a constant but cannot assume that all its derivatives vanish. We are however, going to look at symmetric bouncing solutions, so that near the bounce  we can make the following assumptions about the background quantities:
\be
\begin{split}
R_B&\approx \const\equiv R_b,~~~~\dot R_B\approx 0,~~~~~\Box R_B\approx \const\equiv R_2,\\
H&=0,~~~~~~~~\dot H\approx \const=\frac{R_b}6,~~~~~a\approx 1,~~~~~\dot a=0
\end{split}
\ee
where the subscript $b$ stands for the bounce and numbers indicate the number of derivatives. Under these assumptions
\begin{equation}
\zeta\approx-2R_2\Phi+(\Box_B-r_1)\delta\!R_\text{GI}\label{eqbzeta}
\end{equation}
and equations (\ref{deltatraceGInonconst}) and (\ref{deltaGI1lambdanopis}) simplify to
\ba
(2R_2\Zc(\Box)+\Wc(\Box)|_b)\zeta&=&0\label{eqb11}\\
(\Box_B-r_1)\Zc(\Box)\zeta+\Fc_1[\delta R_\text{GI}+(R_b+3r_1)(\Phi-\Psi)]&=&0
\label{eqb22}
\ea
Note that even if $R_2=0$ equations do not become identical to the ones in the previous section because non-local operators become slightly different due to the new constant $R_b$ instead of $R_\text{dS}$.
It is more difficult do disentangle the above equations but on the other hand the d'Alembertian operator is simpler, thanks to the above mentioned assumptions:
\be
\Box\approx-\pd_t^2-k^2
\ee
and $\delta R_{\text{GI}}$ becomes a bit shorter
\be
\delta\!R_\text{GI}=6\Box_B\Psi-2R_b\Phi+2{k^2}(\Phi+\Psi)
\ee
As the first step we solve the homogeneous equation for $\zeta$ (\ref{eqb11}). Then we express $\Phi$ from (\ref{eqb22}) since the only differential operator acts on $\Psi$ there, and as the last step substitute everything in (\ref{eqbzeta}).
The result will be an equation for $\Psi$ of the fourth order. The main question here is whether there are modes blowing up exactly at the bounce.

In what follows in this Section we are going mainly to specialize to the exact solution (\ref{FRW_Hexact}) so that we can use some nice relations among the background quantities. These kinds of relations are not unique to this solution and the results can be easily generalized. Such  general considerations however do not shed more light on the problem and the most important features are already captured in this particular solution.

In order to consider the general bouncing solutions, and not just the hyperbolic cosine bounce,  we must avoid  the specific background relations for particular solutions. Although tedious, it is certainly  possible to do so. Indeed, we just need to keep values $R_b$ and $R_2$ general, although they are always related to each other, thanks to our general ansatz, as
\be
R_2=r_1 R_b+r_2
\ee
We stress that the parameters $r_1$ and $r_2$ are not initially constrained. The only reason why we are not elaborating on this general case  is that this would significantly complicate all the formulae and would not provide any new insights into the problem. There is nothing special about the hyperbolic cosine bounce except the fact that it has a closed analytic form and also
obeys some very elegant relations among its parameters.

\subsection{Trivial case, $\za=0$}

The simplest solution to Eq.~(\ref{eqb11})  is $\zeta=0$. This  simplifies  Eq.~(\ref{eqb22}) to
\be
6\Box_B\Psi-2R_b\Phi+2{k^2}(\Phi+\Psi)+(R_b+3r_1)(\Phi-\Psi)=0
\label{eqb220}
\ee
and we can deduce $\Phi$ from here as
\be
6\Box_B\Psi+2{k^2}\Psi-(R_b+3r_1)\Psi=
(R_b-3r_1-2k^2)\Phi
\label{eqb220Phi}
\ee
It is important to note that from Section~2.3, see Eq.~(\ref{FRW_Hexact}), $R_b=3r_1$, and therefore equations can be separated in the small $k^2$ limit.
Specializing to the known exact solution we have for $k^2=0$
\be
\ddot\Psi+r_1\Psi=0
\ee
Having $r_1>0$ we get bounded oscillations across the bounce.
Then Eq.~(\ref{eqb11}) reads
\be
\pd_t^2\Phi+2r_1\Phi=0
\ee
where we have used $R_2=-3r_1^2$.
These are again two bounded oscillating modes.

If $k^2>0$ things are more messy. Again discussing only the exact solution from section 2.3, we have
\be
6\ddot\Psi+4{k^2}\Psi+6r_1\Psi=
2k^2\Phi
\label{eqb22k2Phi}
\ee
Using $\zeta$ given by Eq.~(\ref{eqbzeta}) we yield for $\zeta=0$:
\be
r_1({\ddot\Psi+(\frac23{k^2}+r_1)\Psi})+(-\pd^2_t-k^2-r_1)(\frac13k^2\Psi-({\ddot\Psi+(\frac23{k^2}+r_1)\Psi}))=0
\ee
This equation has four solutions
\be
\Psi=c_{\pm\pm}e^{\pm\nu_\pm t},~~~~~~~~\nu_\pm=\frac1{\sqrt{6}}\sqrt{-(9r_1+4k^2)\pm\sqrt{9r_1^2+4k^4}}
\ee
where signs are independent.

We see that all the modes pass the bounce point smoothly and in the limit $k\to0$, modes for $\Phi$ and $\Psi$ would mix together giving the above obtained solutions in $k=0$ case.
What is reassuring is to note that the short wavelength fluctuations ($k\ra \infty$) makes the solutions more and more oscillatory rather than contributing to any exponential growth of fluctuations. This is an encouraging sign in terms of whether these kinds of theories can lead to a truly UV complete theory of gravity.

One may
worry that the right hand side of eq. (\ref{eqb220Phi}) may become zero for a finite value of $k$ for other more general backgrounds. This however, does not change the analysis of the equations and just shifts the point when perturbations decouple to some finite value of the wavenumber $k$.
Finally, we note that these four modes should merge with the GR mode and the $\oa^2=r_1$ mode that was discussed in the previous section during the de Sitter phase.

\subsection{Non-trivial case, $\zeta\neq 0$}
As in the previous section we first solve a simple homogeneous equation of the form
\be
(\Box-\omega_i^2)\zeta\approx-(\pd_t^2+k^2+\omega_i^2)\zeta=0
\ee
This equation has a solution
\be
\zeta_i=z_\pm e^{\pm i\sqrt{k^2+\omega_i^2}t}
\ee
and the exact behaviour depends on the value of $\omega_i$. It is clear, nevertheless, that it passes the bounce point smoothly.
Such a non-zero $\zeta$ serves as an inhomogeneous source term in right hand side of the two remaining equations.

Using Eq.~(\ref{eqb22}), we determine $\Phi$ as
\be
\frac{(\omega_i^2-r_1)}{\Fc_1}\Zc(\omega_i^2)\zeta_i-6(\ddot\Psi+(\frac23{k^2}+r_1)\Psi)=
-2k^2\Phi
\label{eqb22k2zetaPhi}
\ee
where as before we have used particular relations for the exact solution Eq.~(\ref{FRW_Hexact}).
Substituting the results in Eq.~(\ref{eqbzeta}), we get
\be
\begin{split}
r_1({\ddot\Psi+(\frac23{k^2}+r_1)\Psi})+(-\pd^2_t-k^2-r_1)(\frac13k^2\Psi-({\ddot\Psi+(\frac23{k^2}+r_1)\Psi}))-\\
-\frac{\omega_i^2-r_1}{6\Fc_1}(2r_1-\omega_i^2+\frac{k^2}3\frac{\omega_i^2-r_1}{r_1})\Zc(\omega_i^2)\zeta_i=\frac{k^2}{18r_1}\zeta_i
\end{split}
\ee
Extra modes related to the inhomogeneous part will be either oscillatory functions of the frequency determined by $\zeta_i$, or resonance modes if frequencies of $\zeta_i$ and $\Psi$ coincide. We however stress that  even the modes which look dangerous as they may grow exponentially do not blow up while passing the bounce point, and therefore must merge eventually with the late time de Sitter modes discussed in the previous section. They may continue to grow, in which case they must be eliminated according to our criteria (\ref{criteria}), or they must be modes which eventually die out.

\section{Conclusions}\label{sec:conclusion}
In this paper we studied the perturbations of a special class of string-motivated higher derivative extension of Einstein's gravity, which is  well-known to ameliorate the Big Bang cosmological singularity of the FLRW metric. More specifically, our action  contains the Ricci scalar but is non-local in the sense that it contains higher derivative terms up to all orders. It can yield non-singular bouncing solutions similar to that of the {\it hyperbolic cosine } bounce which was first found in~\cite{BMS}. The higher derivatives lead to weakening of the gravity in the UV regime which facilitates the bounce for the FLRW metric in presence of non-ghost like radiation and the cosmological constant.

For the first time we analyzed equations for the scalar perturbations for such non-local higher derivative theories of gravity in complete generality. We provided the general non-local equations that the two  Bardeen potentials satisfy,  and studied the evolution of the generic fluctuations in the relativistic fluid and the scalar potentials during the bounce as well as at late times corresponding to the de Sitter attractor solution. Although we have not included the dynamics and perturbations of the inflaton field, our analysis assessed the robustness of the inflationary scenario in terms of perturbations. We were able to arrive at a simple criteria such that the modes  (small and large $k$ ) pass the bounce phase smoothly, and consequently decay during the de Sitter phase. This means, at least at the classical level there are no pathologies present during the evolution. We also noted the possibility of realizing a curvaton scenario given the extra scalar degree of freedom present in these models are light.

One of the issues we haven't discussed in our paper is  anisotropic deviations from the bouncing background. This requires looking at homogeneous (time dependent) but anisotropic Bianchi I type of models. (The isotropic homogeneous mode was already considered in~\cite{BKM} and issues of stability were discussed.) This is an important issue to determine the robustness of the bouncing mechanism as bouncing/cyclic models are notoriously plagued with Mixmaster type chaotic behavior related to the growth of anisotropies in a contracting universe. The  generalization of the application of the ansatz (\ref{ansatz2}) that has been performed in this paper to arbitrary metrics (potentially inhomogeneous and anisotropic) should help in addressing this issue, but we leave this for future considerations. For stability analysis related to anisotropies in other  non-local gravity and string-inspired models one can see \cite{BI,oobBI} and references therein. It is probably worth pointing out though that anisotropies are expected to be the largest at the bounce point, and thus if there is a ``smooth-enough'' bouncing patch, anisotropies are only going to decrease at large times in the past or future.

For simplicity, some of our analysis was done for the  {\it hyperbolic cosine } bounce, but we discussed how the results should generalize to other bouncing solutions. This strongly suggests that higher derivative actions of gravity considered in this paper do indeed yield a non-singular bounce without a pathology, and in doing so can provide a model of inflation, both in the past and the future.


\acknowledgments
A.K. would like to thank A.Barvinsky, B.Craps and S.Mukhanov for useful and stimulating discussions during the project implementation.
We would also like to thank Nikolay Bulatov for collaborating in the initial phases of this project.

Research of TB is supported by  grants from the Louisiana Board of Regents.
A.K. is the postdoctoral researcher of ``FWO-Vlaanderen'' and is also
supported in part by Belgian Federal Science
Policy Office through the Interuniversity Attraction Pole P6/11, and in
part by the ``FWO-Vlaanderen'' through the project G.0114.10N.
Research of AM is supported by the Lancaster-Manchester-Sheffield Consortium for Fundamental Physics under STFC grant ST/J000418/1.
Research of S.V. is supported in part by  the Russian Ministry of
Education and Science under grant NSh-3920.2012.2, and by contract
CPAN10-PD12 (ICE, Barcelona, Spain).
Research of A.K. and S.V. is supported
in part by the RFBR grant 11-01-00894.

\appendix
\section{Equation for the matter density perturbations}
Scalar stress-energy tensor perturbations are parameterized as follows
\begin{equation*}
T^0_0={}-(\rho+\delta\!\rho),\quad
T^0_i={}-\frac1k(\rho+p)\pd_i v^s,\quad T^i_j=(p+\delta\!
p)\delta^i_j+\left(\frac{\pd^i\pd_j}{k^2}+\frac{\delta^i_j}3\right)\pi^s,
\end{equation*}
where $\rho$ is the energy density, $p$ the pressure, $\delta\rho$ and $\delta p$ are there respective variations, $v^s$ the velocity or the flux related variable and $\pi^s$
the anisotropic stress.  The perturbation functions in $T^\mu_\nu$ are
as follows:
$\delta\!\rho(\eta,x)=\delta\!\rho(\eta,k)e^{ik_jx^j}$ and
similar for $\delta\! p$, $v^s$ and $\pi^s$.
The corresponding often used gauge invariant quantity is~\cite{Mukhanov_rev}-\cite{hwangnoh}:
\begin{equation*}
\varepsilon=\frac{\delta\!\rho}{\rho}+3\frac{aH}{k} \left(1+\frac{p}{\rho}\right)v^s=\frac{\delta\!\rho}{\rho}+3\frac{\Hc}{k} \left(1+\frac{p}{\rho}\right)v^s.
\end{equation*}

 In this paper, we consider only radiation for which
\begin{equation*}
\pi^s=0\,,\qquad  p=\frac{1}{3} \rho\, ,\qquad \delta\! p=\frac{1}{3} \delta\!\rho\, .
\end{equation*}

From the perturbed $(0,0)$ and $(0,i)$ ES equations we express $v^s$ and $\da \rho$ via the Bardeen potentials:
\begin{equation}
\begin{split}
\delta\!\rho=&\lambda\left[\frac{R'_B}{a^2}\Xi'+(r_1 R_B+r_2)\Xi+2\Box_B(\Box_B-r_1)\Xi+2\Fc_1(\zeta+r_1\delta\!R_{\text{GI}})-{}\right.\\
-&4\Fc_1(R_B+3r_1)\left(3\frac{\Hc}{a^2}(\Psi'-\Hc\Phi)+\frac{k^2}{a^2}\Psi\right)
-\left.6\frac{\Hc^2}{a^2}\Upsilon+R_B\Upsilon
+\frac{2}{a^2}\left(\Upsilon''-\Hc \Upsilon'\right)\right]
\end{split}
\label{delta_rho}
\end{equation}
and
\begin{equation}
v^s=\frac{3k\lambda}{2\rho a^2}\left[2\Fc_1(R_B+3r_1)\left(\Psi'-\Hc\Phi\right)
+\frac{1}{2}R'_B\Xi+\Upsilon'-\Hc\Upsilon\right],
\label{deltapreGI0ilambda}
\end{equation}
where
\begin{equation*}
\Upsilon=(\Box_B-r_1)\Xi+\Fc_1\delta\!R_{\text{GI}}\quad\mand\quad \Xi=Z(\Box)\za.
\end{equation*}

The resulting expression for $\varepsilon$ then is
\begin{equation}
\begin{split}
\varepsilon=&{} \frac{\lambda}{\rho}\left[\frac{R'_B}{a^2}\Xi'+(r_1 R_B+r_2)\Xi+2\Box_B(\Box_B-r_1)\Xi+3\frac{\Hc R'_B}{a^2}\Xi+2\Fc_1(\zeta+r_1\delta\!R_{\text{GI}})-{}\right.\\
-&{}\left. 4\frac{k^2}{a^2}\Fc_1(R_B+3r_1)\Psi+\frac{\Hc}{a^2}R'_B\Xi
+\left\{R_B-12\frac{\Hc^2}{a^2}\right\}\Upsilon
+\frac{2}{a^2}\Upsilon''\right].
\end{split}
\end{equation}
One interesting special case is $R_B\to -3r_1$, but this is not a limit for the known background solutions.

\end{document}